\documentclass[aps, prd, amsmath, floats, floatfix, twocolumn, nofootinbib, superscriptaddress, showpacs]{revtex4-1}
\usepackage{graphicx}
\usepackage{color}
\usepackage{latexsym}

\newcommand{\be}{\begin{eqnarray}}
\newcommand{\ee}{\end{eqnarray}}
\newcommand{\rar}{\rightarrow}

\begin{document}

\title{Testing conformal gravity with astrophysical black holes}

\author{Cosimo Bambi}
\email{bambi@fudan.edu.cn}
\affiliation{Center for Field Theory and Particle Physics and Department of Physics, Fudan University, 200433 Shanghai, China}
\affiliation{Theoretical Astrophysics, Eberhard-Karls Universit\"at T\"ubingen, 72076 T\"ubingen, Germany}

\author{Zheng Cao}
\affiliation{Center for Field Theory and Particle Physics and Department of Physics, Fudan University, 200433 Shanghai, China}

\author{Leonardo Modesto}
\email{lmodesto@sustc.edu.cn}
\affiliation{Department of Physics, Southern University of Science and Technology, Shenzhen 518055, China}

\date{\today}

\begin{abstract}
Weyl conformal symmetry can solve the problem the spacetime singularities present in Einstein's gravity. In a recent paper, two of us have found a singularity-free rotating black hole solution in conformal gravity. In addition to the mass $M$ and the spin angular momentum $J$ of the black hole, the new solution has a new parameter, $L$, which here we consider to be proportional to the black hole mass. Since the solution is conformally equivalent to the Kerr metric, photon trajectories are unchanged, while the structure of an accretion disk around a black hole is affected by the value of the parameter $L$. In this paper, we show that X-ray data of astrophysical black holes require $L/M < 1.2$. 
\end{abstract}

\maketitle

%%%%%%%%%%%%%%%%%%%%%%%%%%%%%%%

\section{Introduction}

Einstein's theory of general relativity has been able to pass a large number of tests, and there is currently no clear evidence of disagreement between theoretical predictions and observational data~\cite{will}. However, we know that the theory breaks down in some extreme conditions. In particular, there are physically relevant solutions with spacetime singularities, where predictability is lost and standard physics cannot be applied.

The resolution of spacetime singularities in Einstein's gravity is an outstanding and longstanding problem. Different authors have explored different approaches. It is possible that the resolution of spacetime singularities in Einstein's gravity is related to the yet unknown theory of quantum gravity. Among the many proposals present in the literature, in this paper we are interested in the family of conformal theories of gravity~\cite{cg1,cg2,cg3,cg4,cg5}.

In conformal gravity, the theory is invariant under a conformal transformation of the metric tensor; that is,
\be\label{eq-conf}
g_{\mu\nu} \rar g_{\mu\nu}^* = \Omega^2 g_{\mu\nu} \, ,
\ee
where $\Omega = \Omega (x)$ is a function of the point of the spacetime. Einstein's gravity is not conformally invariant, but it can be made conformally invariant, for instance by introducing an auxiliary field. There are many possible realizations in the literature~\cite{cg1,cg2,cg3,cg4,cg5}. Examples of conformal theories of gravity in four dimensions are
\be
\mathcal{L}_1 &=& a \, C^{\mu\nu\rho\sigma} C_{\mu\nu\rho\sigma} 
+ b \, R^{\mu\nu\rho\sigma} R_{\mu\nu\rho\sigma} \, , \nonumber\\
\mathcal{L}_2 &=& \phi^2 R + 6 \, g^{\mu\nu} (\partial_\mu \phi)( \partial_\nu \phi) \, , \nonumber
\ee
where $C^{\mu\nu\rho\sigma}$ is the Weyl tensor, $R^{\mu\nu\rho\sigma}$ is the Riemann tensor, $a$ and $b$ are constants, and $\phi$ is an auxiliary scalar field (dilaton). In our case, we are not interested in a particular model, but we just require the theory to be invariant under the transformation~(\ref{eq-conf}). While the theory is invariant under conformal transformations, a Higgs-like mechanism may choose one of the metric as the ``physical'' solution to describe the spacetime. The Universe indeed does not appear to be conformally invariant, but there is evidence of scale invariance in the early Universe and in many other physical phenomena, like phase transitions, etc. Moreover, we tend to believe that the physics very close to the black hole may be described by a quantum gravity theory in its conformal invariant phase because the Hawking temperature gets a huge blueshift near the event horizon. We could thus expect large deviations from the Kerr metric because the energy becomes very large (trans Planckian) and the conformal invariance is restored while the vacuum becomes  degenerate.

Conformal gravity can solve the problem of spacetime singularities by finding a suitable conformal transformation $\Omega$ that removes the singularity and by interpreting the metric $g_{\mu\nu}^*$ as the physical metric of the spacetime. In Ref.~\cite{metric}, we found a singularity-free rotating black hole solution conformally equivalent to the Kerr metric (see Ref.~\cite{m-q} on how conformal invariance is preserved at the quantum level). In Boyer-Lindquist coordinates, the line element reads
\be\label{eq-m}
ds^2 = \left(1 + \frac{L^2}{\Sigma}\right)^4 ds^2_{\rm Kerr} \, ,
\ee
where $L > 0$ is a new parameter, $\Sigma = r^2 + a^2 \cos^2\theta$, and $ds^2_{\rm Kerr}$ is the line element of the Kerr metric
\be
&& \hspace{-0.1cm}
ds^2_{\rm Kerr} = - \left(1 - \frac{2 M r}{\Sigma}\right) dt^2 
- \frac{4 M a r \sin^2\theta}{\Sigma} dt d\phi \\
&& \hspace{-0.1cm}
+ \frac{\Sigma}{\Delta} dr^2 
+ \Sigma d\theta^2  + \left(r^2 + a^2 + \frac{2 M a^2 r \sin^2\theta}{\Sigma} \right) 
\sin^2\theta d\phi^2 \, , \nonumber 
\ee
$a = J/M$ is the rotational parameter (the dimensionless spin parameter is $a_* = a/M$) and $\Delta = r^2 - 2 M r + a^2$. Let us note that Eq.~(\ref{eq-m}) is an exact rotating black hole solution in a large family of theories. Black holes in theoretically-motivated alternative theories of gravity are usually known in the non-rotating case, sometimes in the slow-rotation approximation, but there are only a few examples in the literature of exact solutions for any value of the spin.

Black holes in theories beyond Einstein's gravity have been extensively investigated in the past 10-20~years. They may have different theoretical (e.g. thermodynamics stability, uniqueness of solutions, entropy, topology of the horizon, etc.) and observational properties with respect to the Kerr black holes of Einstein's gravity, so that well-known results valid in the standard theory may not hold. See, for instance, \cite{ppp0,ppp1,ppp2,Berti:2015itd,review} and reference therein.

In this work, we want to ``test'' the metric in Eq.~(\ref{eq-m}) and constrain the parameter $L$. At the moment there are no indications from the theory about the value of $L$, which may thus be expected either of the order of 1 (in Planck units) or of the order of $M$. There are no other scales. If $L$ is of the order of the Planck length, there is no way to get an estimate of its value with the technique discussed in this paper. We thus consider the second, more favorable, case, in which $L$ can be of the order of $M$. There are indeed scenarios in the literature in which one can expect macroscopic deviations at the scale of the horizon~\cite{q1,q2,q3,q4}. As we will show below within a simple analysis, current X-ray data of astrophysical black holes require $L/M < 1.2$. Since the metric in Eq.~(\ref{eq-m}) is conformally equivalent to the Kerr solution, photon trajectories are the same as in the Kerr metric. However, the structure of the accretion disk changes. In particular, the value of $L$ alters the innermost stable circular orbit (ISCO), which has a strong impact on current techniques like the continuum-fitting and the iron line methods.

\begin{figure}[t]
%\vspace{0.5cm}
\begin{center}
\includegraphics[type=pdf,ext=.pdf,read=.pdf,width=8cm]{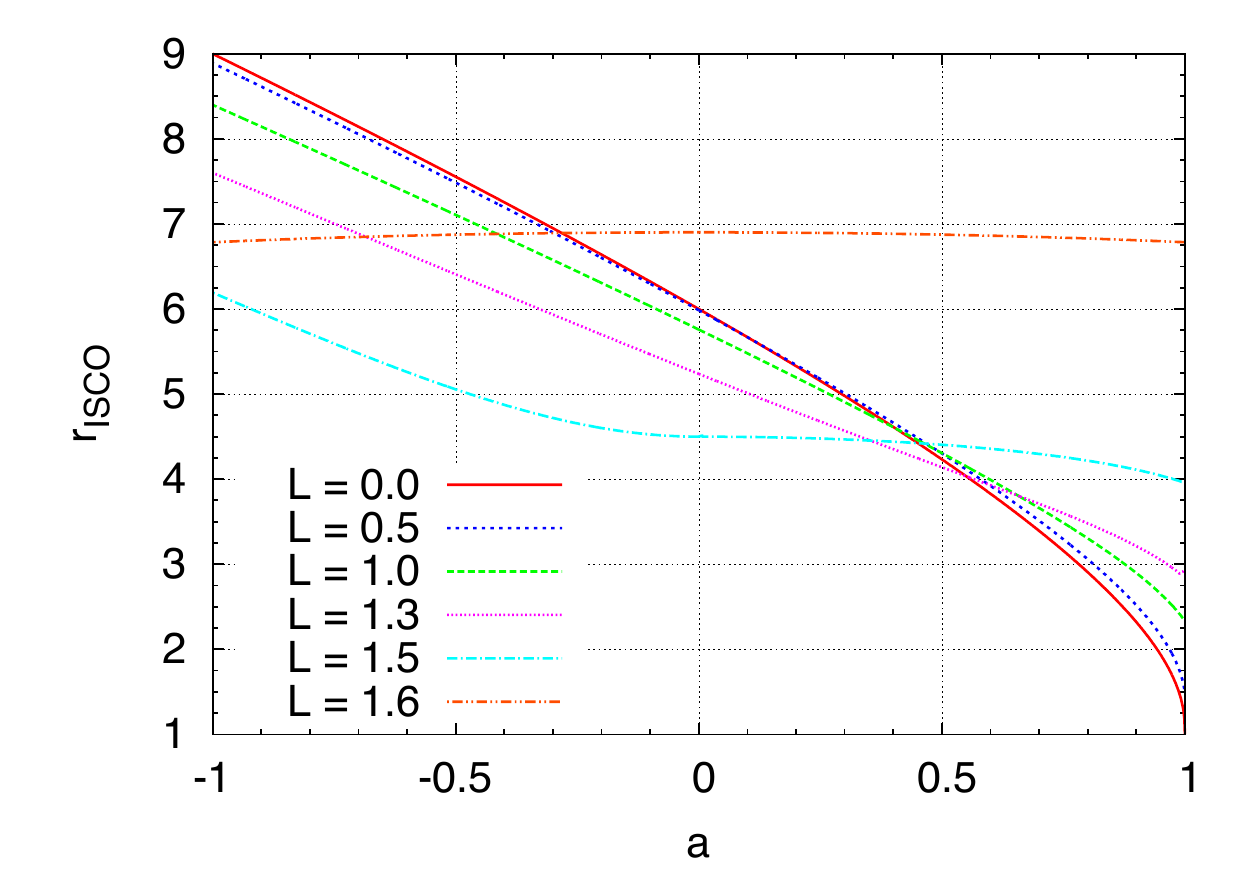}
\end{center}
\vspace{-0.4cm}
\caption{ISCO radius $r_{\rm ISCO}$ as a function of the specific spin $a$ for different values of the parameter $L$. $r_{\rm ISCO}$, $a$, and $L$ in units in which $M=1$. \label{f-isco}}
\end{figure}

\begin{figure*}[t]
%\vspace{0.5cm}
\begin{center}
\includegraphics[type=pdf,ext=.pdf,read=.pdf,width=8.5cm]{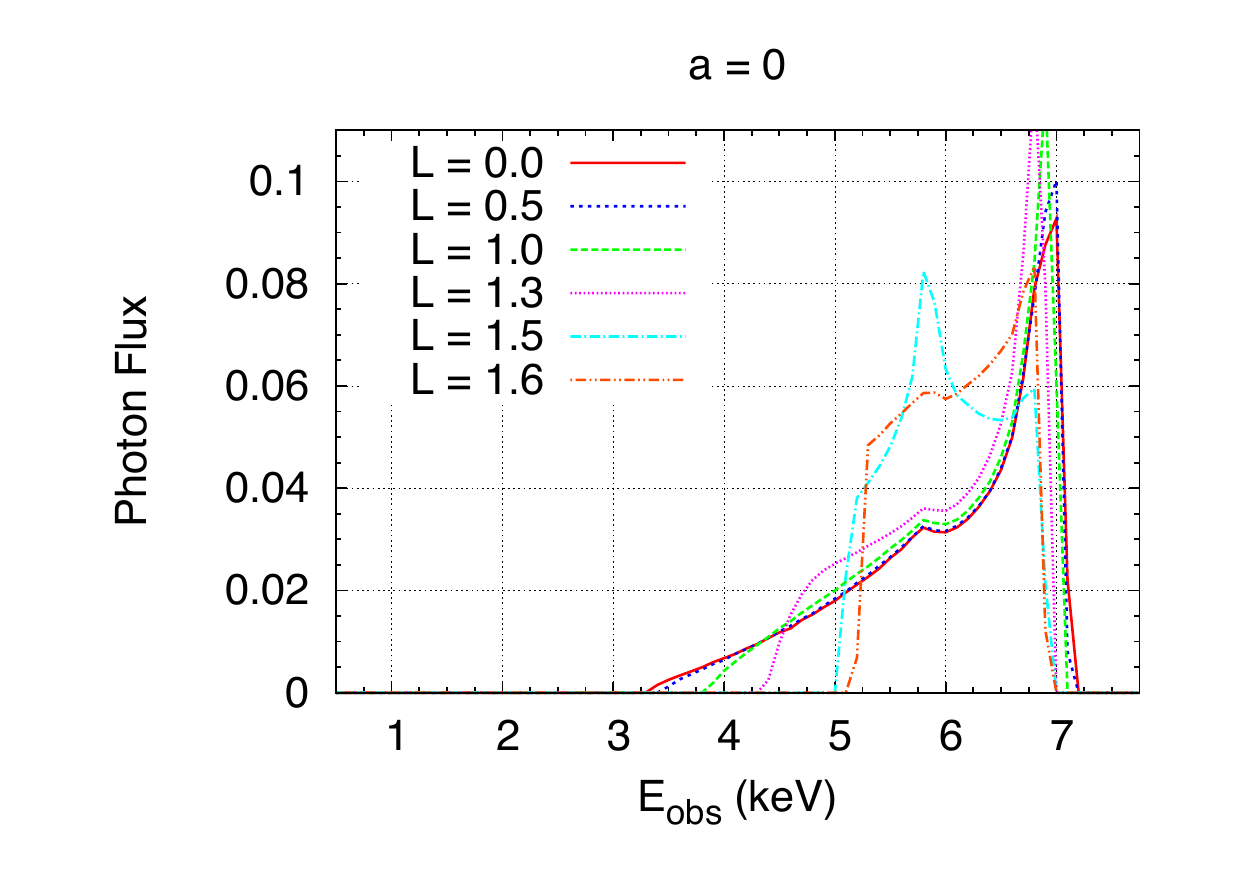}
\includegraphics[type=pdf,ext=.pdf,read=.pdf,width=8.5cm]{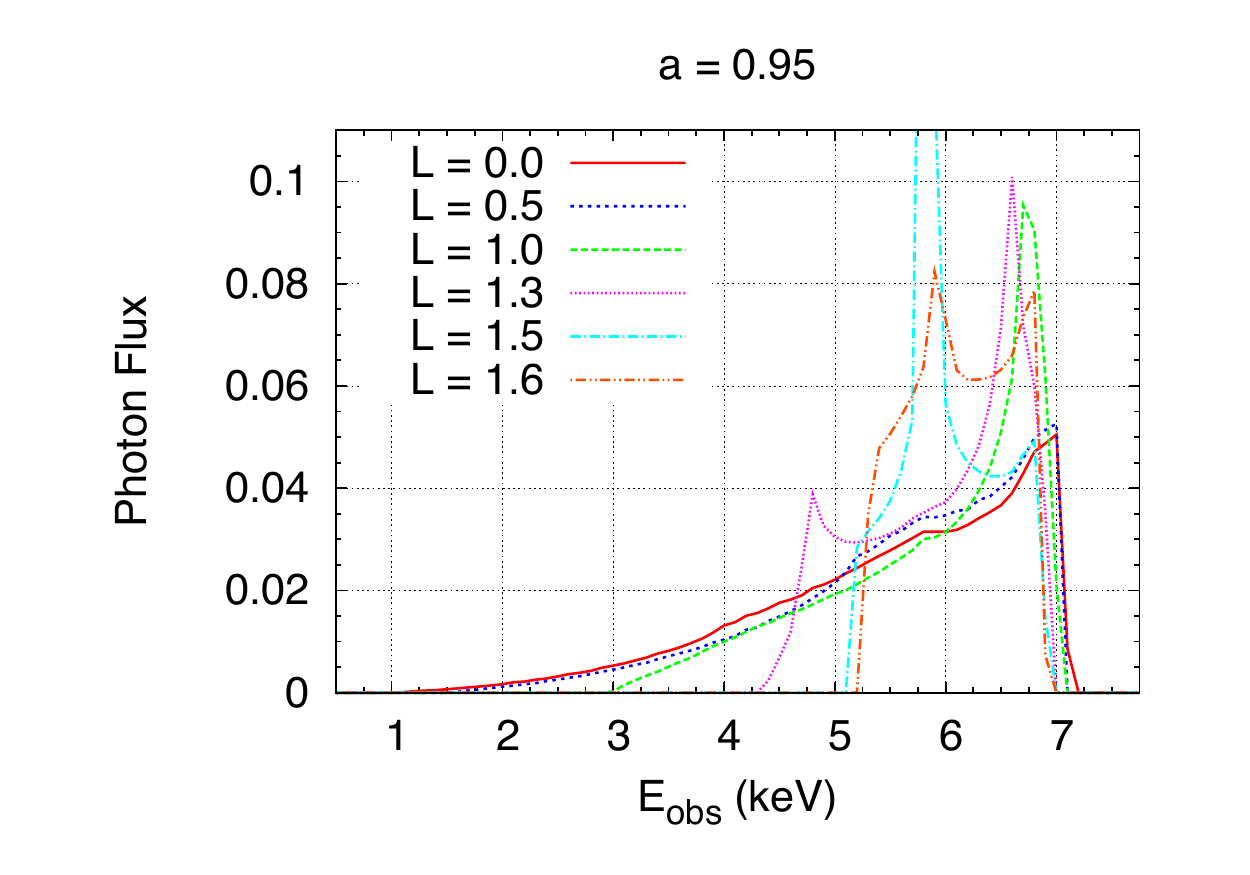}
\end{center}
\vspace{-0.6cm}
\caption{Iron line shapes in the case of non-rotating black hole solutions ($a_* = 0$, left panel) and fast-rotating black hole solutions ($a_* = 0.95$, right panel) for different values of the conformal length $L$. The viewing angle in these simulations is $i = 45^\circ$ and the emissivity profile of the disk is assumed $\propto 1/r^3$. \label{f-line}}
\end{figure*}

\section{Accretion disk}

There are a number of proposals to probe the spacetime metric around astrophysical black holes with electromagnetic radiation~\cite{review,review2}. At present, the continuum-fitting and the iron line methods can provide some general constraints, while other techniques are not yet mature to test fundamental physics or the necessary observational data are not yet available.

The continuum-fitting and the iron line methods assume that the accretion disk is described by the Novikov-Thorne model~\cite{nt1,nt2}, which is the standard framework for geometrically thin and optically thick accretion disks around black holes. The disk is in the equatorial plane, perpendicular to the black hole spin. The particles of the gas follow nearly geodesic equatorial circular orbits. A crucial ingredient of the model is that the inner edge of the disk is at the ISCO radius. This assumption plays a fundamental role in the continuum-fitting and the iron line methods and is supported by the observed stability of the position of the inner edge of the disk~\cite{jack}.

The ISCO radius for a generic stationary, axisymmetric, and asymptotically flat spacetime can be computed as follows (see, for instance, Appendix~B in~\cite{enrico} for more details). We write the line element in the canonical form, namely
\be
\hspace{-0.4cm}
ds^2 = g_{tt} dt^2 + g_{t\phi} dt d\phi + g_{rr} dr^2 
+ g_{\theta\theta} d\theta^2 + g_{\phi\phi} d\phi^2 \, ,
\ee
where the metric coefficients are independent of $t$ and $\phi$. We have thus two constants of motion, namely the specific energy $E$ and the axial component of the specific angular momentum $L_z$. We write the $t$- and $\phi$-components of the 4-velocity of a particle in terms of $E$, $L_z$, and the metric coefficients. From the conservation of the rest-mass $g_{\mu\nu} \dot{x}^\mu \dot{x}^\nu = - 1$, we write
\be
g_{rr} \dot{r}^2 + g_{\theta\theta} \dot{\theta}^2 = V_{\rm eff} (r, \theta) \, ,
\ee
where the dot $\dot{}$ indicates the derivative with respect to the particle proper time and the effective potential $V_{\rm eff}$ is
\be
V_{\rm eff} = \frac{E^2 g_{\phi\phi} + 2 E L_z g_{t\phi} 
+ L_z^2 g_{tt}}{g_{t\phi}^2 - g_{tt}g_{\phi\phi}} - 1 \, .
\ee
Circular orbits in the equatorial plane have $V_{\rm eff} = \partial_r V_{\rm eff} = \partial_\theta V_{\rm eff} = 0$. The orbits are stable under small perturbations if $\partial^2_r V_{\rm eff} \le 0$ and $\partial^2_\theta V_{\rm eff} \le 0$. The ISCO radius is found when either $\partial^2_r V_{\rm eff}$ or $\partial^2_\theta V_{\rm eff}$ vanish.

Fig.~\ref{f-isco} shows the ISCO radius $r_{\rm ISCO}$ as a function of the rotational parameter $a$ for different values of $L$. The value of the radial coordinate has not a direct physical meaning, as it depends on the choice of the coordinates, which is arbitrary. However, from Fig.~\ref{f-isco} we see that, as the value of $L$ increases, the minimum value of $r_{\rm ISCO}$ increases too. We can already anticipate that this implies that very broad iron lines -- as we observe in the X-ray data of some black holes -- are not possible for large $L$, because the gravitational redshift would be too weak.

\section{Iron K$\alpha$ line}

Now we want to take a step forward and calculate the iron K$\alpha$ line that can be expected in the reflection spectrum of an accretion disk around a black hole with $L > 0$. Within the disk-corona model~\cite{corona1,corona2}, we have a thin accretion disk surrounding a black hole. The corona is a hot ($\sim 100$~keV), usually optically thin, cloud. The corona may be the base of a jet, a sort of atmosphere above the accretion disk, etc. The actual geometry is currently unknown. Due to inverse Compton scattering of thermal photons from the disk from free electrons in the corona, the latter becomes an X-ray source with a power-law spectrum. A fraction of these X-ray photons can illuminate the disk, producing the so-called reflection spectrum with some fluorescent emission lines~\cite{refl}. The most prominent line is usually the iron K$\alpha$ one, which is at 6.4~keV in the case of neutral iron and shifts up to 6.97~keV in the case of H-like iron ions.

The iron K$\alpha$ line is very narrow in the rest-frame of the emitter, but it can appear broad and skewed in the spectrum of a black hole as the result of relativistic effects (Doppler boosting, gravitational redshift, light bending) occurring in the strong gravity region. If we assume the Kerr metric, the analysis of the iron line can provide an estimate of the black hole spin~\cite{iron1,iron2}. If we relax the Kerr black hole hypothesis, this technique can be used to constrain possible deviations from the Kerr solution. Let us note that, in the presence of high quality data and of the correct astrophysical model, the iron line method can be a powerful tool to test the spacetime metric around astrophysical black holes~\cite{icb1,icb2,icb3}. Actually, one has to fit the whole reflection spectrum, not just the iron line, but the latter is the most prominent feature and, in a preliminary analysis like the present work, we can restrict our attention to the iron K$\alpha$ line only.

Fig.~\ref{f-line} shows the expected iron line profiles in the reflection spectrum of a black hole with, respectively, the spin parameter $a_* = 0$ (left panel) and 0.95 (right panel) and different values of $L$. The calculations have been done with the code described in Refs.~\cite{c1,c2}. The case $L=0$ corresponds to the Kerr metric. In these calculations, we have employed the emissivity profile $\propto 1/r^3$, which corresponds to the case of the Newtonian limit (no light bending) at large radii for a point-like corona just above the black hole. However, for our considerations it does not play an important role, because the choice of the intensity profile can only alter the shape of the iron line. It cannot determine the photon energy detected at infinity from a specific point of the disk. From Fig.~\ref{f-line}, we see that, as $L$ increases, the observed iron line profile becomes less broad. This is true even for the non-rotating case (left panel) although the ISCO radius may actually be smaller than the case with $L=0$ (see Fig.~\ref{f-isco}). X-ray data of some black holes clearly show very broad iron line that can extend to low energies. This would not be possible with a sufficiently high value of $L$, and this permits us to constrain the value of $L$.

\section{Simulations}

Let us now be more quantitative and get a constraint on the value of $L$ from the iron line. A rough estimate of the maximum value of $L$ allowed by current X-ray data can be obtained quite easily following the method already employed in Refs.~\cite{p1,p2,p3,p4}. We know that current X-ray data are consistent with the Kerr metric, in the sense that observations are fitted with theoretical Kerr model and we obtain acceptable fits. We also know that some iron line are very broad and, if interpreted within the Kerr metric, it means that the spin parameter of the black hole is close to 1. For instance, assuming the Kerr metric, for Cygnus~X-1 we have the measurement $a_* = 0.97^{+0.014}_{-0.02}$~\cite{fast1}, for LMC~X-1 $a_* = 0.97^{+ 0.01}_{-0.13}$~\cite{fast2} , for GX~339-4 $a_* = 0.95^{+ 0.03}_{-0.05}$~\cite{fast3}.

As done in Refs.~\cite{p1,p2,p3,p4} for other scenarios, we simulate some observations taking into account the response of the instrument, the background noise, and the intrinsic Poisson noise of the source. We employ the typical parameters for a bright stellar-mass black hole in a binary. We model the X-ray spectrum of the source with a power-law $\propto E^{-\Gamma}$ with $\Gamma = 2$ (the primary spectrum of the corona) and a single iron line (the reflection spectrum from the disk). This is a simple model, but it should capture the main features of the new metric and is enough for getting a rough constraint on $L$. We assume that the energy flux in the 3-10~keV range is $6 \cdot 10^{-9}$~erg/s/cm$^2$ and that the equivalent width of the iron line is around 200~eV. We simulate observation with NuSTAR\footnote{http://www.nustar.caltech.edu}, assuming that the exposure time is 100~ks.

We have simulated a number of observations for $a_* = 0.95$ and different values of the parameters $L$, of the viewing angle $i$, and with different emissivity profiles. The simulations have been treated as real data and fitted with XSPEC\footnote{XSPEC is an X-ray spectral-fitting software commonly used in X-ray astronomy. See~\cite{arn96} and http://heasarc.gsfc.nasa.gov/docs/xanadu/xspec/index.html for more details.}, employing the model RELLINE for the iron line~\cite{relline}. If $L$ is not very large, the fits are usually acceptable, in the sense that there are not unresolved features and the minimum of the reduced $\chi^2$ is not much higher than 1. If we require that the analysis of our simulations should provide spin measurements $a_* > 0.9$, we find that a conservative constraint for $L$ is $L/M < 1.2$. More stringent constraints can be definitively obtained with a more detailed analysis, which is beyond the explorative work presented in this work.

\section{Concluding remarks}

In this paper, we have discussed how to test conformal gravity with observational data of astrophysical black holes. In particular, we have considered the singularity-free rotating black hole solution in Eq.~(\ref{eq-m}) and obtained a rough constraint on the parameter $L$. The latter can change the structure of the accretion disk around black holes and, in particular, can alter the ISCO radius, a quantity that can have a strong impact on the X-ray spectrum of black holes. Within a very simple analysis, we have discussed the expected iron line in the reflection spectrum of these spacetimes and shown that current observations require $L/M < 1.2$. For higher values of $L$, it is not possible to reproduce the extended low energy tail in the iron line as observed in the spectra of several sources.

We would like to remark that our analysis is quite simple, but the constraint $L/M < 1.2$ is robust and is not affected by our assumptions on the astrophysical model. In particular, in typical spin measurements or tests of the Kerr metric, a crucial point is the emissivity profile, namely the flux of the reflection spectrum at different radii. Changing the emissivity profile, we alter the relative contribution from different parts of the accretion disk. Our result does not depend on the intensity profile, because the key-point of the constraint $L/M < 1.2$ is that observations show broad iron lines with a low energy tail up to $\sim 1$~keV, while for sufficiently high values of $L/M$ we cannot have very redshifted photons at any radius of the accretion disk. The photon redshift is only determined by the background metric, and it is not possible to increase the redshift factor of the photons by changing the astrophysical model. A more detailed analysis with the more sophisticated model presented in Ref.~\cite{next} can study real data of specific sources and get stronger constraints. We postpone such an analysis to a future work.

%%%%%%%%%%%%%%%%%%%%%%%%%%%%%%%

\begin{acknowledgments}
This work was supported by the NSFC (grants U1531117 and 11305038) and the Thousand Young Talents Program. C.B. also acknowledges the support from the Alexander von Humboldt Foundation.
\end{acknowledgments}

\pagebreak

%%%%%%%%%%%%%%%%%%%%%%%%%%%%%%%


\begin{thebibliography}{99}

\bibitem{will} 
  C.~M.~Will,
  %``The Confrontation between General Relativity and Experiment,''
  Living Rev.\ Rel.\  {\bf 17}, 4 (2014)
  [arXiv:1403.7377 [gr-qc]].

\bibitem{cg1} 
  F.~Englert, C.~Truffin and R.~Gastmans,
  %``Conformal Invariance in Quantum Gravity,''
  Nucl.\ Phys.\ B {\bf 117}, 407 (1976).

\bibitem{cg2} 
  J.~V.~Narlikar and A.~K.~Kembhavi,
  %``Space-Time Singularities and Conformal Gravity,''
  Lett.\ Nuovo Cim.\  {\bf 19}, 517 (1977).

\bibitem{cg3} 
  G.~'t Hooft,
  %``Quantum gravity without space-time singularities or horizons,''
  Subnucl.\ Ser.\  {\bf 47}, 251 (2011)
  [arXiv:0909.3426 [gr-qc]].

\bibitem{cg4} 
  P.~D.~Mannheim,
  %``Making the Case for Conformal Gravity,''
  Found.\ Phys.\  {\bf 42}, 388 (2012)
  [arXiv:1101.2186 [hep-th]].
  
\bibitem{cg5} 
  I.~Bars, P.~Steinhardt and N.~Turok,
  %``Local Conformal Symmetry in Physics and Cosmology,''
  Phys.\ Rev.\ D {\bf 89}, 043515 (2014)
  [arXiv:1307.1848 [hep-th]].

\bibitem{metric} 
  C.~Bambi, L.~Modesto and L.~Rachwal,
  %``Spacetime completeness of non-singular black holes in conformal gravity,''
  arXiv:1611.00865 [gr-qc].
  
\bibitem{m-q} 
  L.~Modesto and L.~Rachwal,
  %``Super-renormalizable and finite gravitational theories,''
  Nucl.\ Phys.\ B {\bf 889}, 228 (2014)
  [arXiv:1407.8036 [hep-th]].  
  
\bibitem{ppp0} 
  S.~Nojiri and S.~D.~Odintsov,
  %``Anti-de Sitter black hole thermodynamics in higher derivative gravity and new confining deconfining phases in dual CFT,''
  Phys.\ Lett.\ B {\bf 521}, 87 (2001)
  [Erratum: ibidem\ {\bf 542}, 301 (2002)]
  [hep-th/0109122].  
  
\bibitem{ppp1} 
  M.~Cvetic, S.~Nojiri and S.~D.~Odintsov,
  %``Black hole thermodynamics and negative entropy in de Sitter and anti-de Sitter Einstein-Gauss-Bonnet gravity,''
  Nucl.\ Phys.\ B {\bf 628}, 295 (2002)
  [hep-th/0112045].

\bibitem{ppp2} 
  A.~de la Cruz-Dombriz and D.~Saez-Gomez,
  %``Black holes, cosmological solutions, future singularities, and their thermodynamical properties in modified gravity theories,''
  Entropy {\bf 14}, 1717 (2012)
  [arXiv:1207.2663 [gr-qc]].  
  
\bibitem{Berti:2015itd} 
  E.~Berti {\it et al.},
  %``Testing General Relativity with Present and Future Astrophysical Observations,''
  Class.\ Quant.\ Grav.\  {\bf 32}, 243001 (2015)
  [arXiv:1501.07274 [gr-qc]].  
  
\bibitem{review} 
  C.~Bambi,
  %``Testing black hole candidates with electromagnetic radiation,''
  Rev.\ Mod.\ Phys.\  (in press)
  [arXiv:1509.03884 [gr-qc]].  

\bibitem{q1} 
  S.~D.~Mathur,
  %``The Fuzzball proposal for black holes: An Elementary review,''
  Fortsch.\ Phys.\  {\bf 53}, 793 (2005)
  [hep-th/0502050].

\bibitem{q2} 
  G.~Dvali and C.~Gomez,
  %``Black Hole's Quantum N-Portrait,''
  Fortsch.\ Phys.\  {\bf 61}, 742 (2013)
  [arXiv:1112.3359 [hep-th]].  
  
\bibitem{q3} 
  G.~Dvali and C.~Gomez,
  %``Black Hole's 1/N Hair,''
  Phys.\ Lett.\ B {\bf 719}, 419 (2013)
  [arXiv:1203.6575 [hep-th]].  
  
\bibitem{q4} 
  S.~B.~Giddings,
  %``Possible observational windows for quantum effects from black holes,''
  Phys.\ Rev.\ D {\bf 90}, 124033 (2014)
  [arXiv:1406.7001 [hep-th]].  
  
\bibitem{review2} 
  C.~Bambi, J.~Jiang and J.~F.~Steiner,
  %``Testing the no-hair theorem with the continuum-fitting and the iron line methods: a short review,''
  Class.\ Quant.\ Grav.\  {\bf 33}, 064001 (2016)
  [arXiv:1511.07587 [gr-qc]].  
  
\bibitem{nt1}
  I.~D.~Novikov and K.~S.~Thorne,
  {\it Astrophysics and black holes}, in {\it Black Holes}, edited by C.~De~Witt and B.~De~Witt(Gordon and Breach, New York, US, 1973).

\bibitem{nt2} 
  D.~N.~Page and K.~S.~Thorne,
  %``Disk-Accretion onto a Black Hole. Time-Averaged Structure of Accretion Disk,''
  Astrophys.\ J.\  {\bf 191}, 499 (1974).  
  
\bibitem{jack} 
  J.~F.~Steiner, J.~E.~McClintock, R.~A.~Remillard, L.~Gou, S.~Yamada and R.~Narayan,
  %``The Constant Inner-Disk Radius of LMC X-3: A Basis for Measuring Black Hole Spin,''
  Astrophys.\ J.\  {\bf 718}, L117 (2010)
  [arXiv:1006.5729 [astro-ph.HE]].    
  
\bibitem{enrico} 
  C.~Bambi and E.~Barausse,
  %``Constraining the quadrupole moment of stellar-mass black-hole candidates with the continuum fitting method,''
  Astrophys.\ J.\  {\bf 731}, 121 (2011)
  [arXiv:1012.2007 [gr-qc]].  
  
\bibitem{corona1} 
  G.~Matt, G.~C.~Perola and L.~Piro,
  Astron.\ Astrophys.\  {\bf 247}, 25 (1991).

\bibitem{corona2} 
  A.~Martocchia and G.~Matt,
  Mon.\ Not.\ Roy.\ Astron.\ Soc.\  {\bf 282}, L53 (1996).  
  
\bibitem{refl} 
  A.~C.~Fabian, K.~Iwasawa, C.~S.~Reynolds and A.~J.~Young,
  %``Broad iron lines in active galactic nuclei,''
  Publ.\ Astron.\ Soc.\ Pac.\  {\bf 112}, 1145 (2000)
  [astro-ph/0004366].  
  
\bibitem{iron1} 
  L.~W.~Brenneman and C.~S.~Reynolds,
  %``Constraining Black Hole Spin Via X-ray Spectroscopy,''
  Astrophys.\ J.\  {\bf 652}, 1028 (2006)
  [astro-ph/0608502].  
  
\bibitem{iron2} 
  C.~S.~Reynolds,
  %``Measuring Black Hole Spin using X-ray Reflection Spectroscopy,''
  Space Sci.\ Rev.\  {\bf 183}, 277 (2014)
  [arXiv:1302.3260 [astro-ph.HE]].    
  
\bibitem{icb1} 
  J.~Jiang, C.~Bambi and J.~F.~Steiner,
  %``Testing the Kerr Nature of Black Hole Candidates using Iron Line Spectra in the CPR Framework,''
  Astrophys.\ J.\  {\bf 811}, 130 (2015)
  [arXiv:1504.01970 [gr-qc]].  
  
\bibitem{icb2} 
  A.~Cardenas-Avendano, J.~Jiang and C.~Bambi,
  %``Testing the Kerr black hole hypothesis: comparison between the gravitational wave and the iron line approaches,''
  Phys.\ Lett.\ B {\bf 760}, 254 (2016)
  [arXiv:1603.04720 [gr-qc]].  
  
\bibitem{icb3} 
  Y.~Ni, J.~Jiang and C.~Bambi,
  %``Testing the Kerr metric with the iron line and the KRZ parametrization,''
  JCAP {\bf 1609}, 014 (2016)
  [arXiv:1607.04893 [gr-qc]].  
    
\bibitem{c1} 
  C.~Bambi,
  %``A code to compute the emission of thin accretion disks in non-Kerr space-times and test the nature of black hole candidates,''
  Astrophys.\ J.\  {\bf 761}, 174 (2012)
  [arXiv:1210.5679 [gr-qc]].  
  
\bibitem{c2} 
  C.~Bambi,
  %``Testing the space-time geometry around black hole candidates with the analysis of the broad K$\alpha$ iron line,''
  Phys.\ Rev.\ D {\bf 87}, 023007 (2013)
  [arXiv:1211.2513 [gr-qc]].  
  
\bibitem{p1} 
  M.~Zhou, A.~Cardenas-Avendano, C.~Bambi, B.~Kleihaus and J.~Kunz,
  %``Search for astrophysical rotating Ellis wormholes with X-ray reflection spectroscopy,''
  Phys.\ Rev.\ D {\bf 94}, 024036 (2016)
  [arXiv:1603.07448 [gr-qc]].  
  
\bibitem{p2} 
  Y.~Ni, M.~Zhou, A.~Cardenas-Avendano, C.~Bambi, C.~A.~R.~Herdeiro and E.~Radu,
  %``Iron K$\alpha$ line of Kerr black holes with scalar hair,''
  JCAP {\bf 1607}, 049 (2016)
  [arXiv:1606.04654 [gr-qc]].  
  
\bibitem{p3} 
  Z.~Cao, A.~Cardenas-Avendano, M.~Zhou, C.~Bambi, C.~A.~R.~Herdeiro and E.~Radu,
  %``Iron K$\alpha$ line of boson stars,''
  JCAP {\bf 1610}, 003 (2016)
  [arXiv:1609.00901 [gr-qc]].  
  
\bibitem{p4} 
  M.~Ghasemi-Nodehi and C.~Bambi,
  %``Constraining the Kerr parameters via X-ray reflection spectroscopy,''
  Phys.\ Rev.\ D {\bf 94}, 104062 (2016)
  [arXiv:1610.08791 [gr-qc]].  
  
\bibitem{fast1}  
  A.~C.~Fabian {\it et al.},
  %``On the determination of the spin of the black hole in Cyg X-1 from X-ray reflection spectra,''
  Mon.\ Not.\ Roy.\ Astron.\ Soc.\  {\bf 424}, 217 (2012)
  [arXiv:1204.5854 [astro-ph.HE]].   
  
\bibitem{fast2}   
  J.~F.~Steiner {\it et al.},
  %``A Broad Iron Line in LMC X-1,''
  Mon.\ Not.\ Roy.\ Astron.\ Soc.\  {\bf 427}, 2552 (2012)
  [arXiv:1209.3269 [astro-ph.HE]].   
  
\bibitem{fast3} 
  J.~Garcia {\it et al.},
  %``X-Ray Reflection Spectroscopy of the Black Hole GX 339--4: Exploring the Hard State with Unprecedented Sensitivity,''
  Astrophys.\ J.\ {\bf 813}, 84 (2015)
  [arXiv:1505.03607[astro-ph.HE]].    
  
\bibitem{arn96}  
  K.~A.~Arnaud,
  Astronomical Data Analysis Software and Systems V, {\bf 101}, 17 (1996).  
  
\bibitem{relline} 
  T.~Dauser, J.~Wilms, C.~S.~Reynolds and L.~W.~Brenneman,
  %``Broad emission lines for negatively spinning black holes,''
  Mon.\ Not.\ Roy.\ Astron.\ Soc.\  {\bf 409}, 1534 (2010)
  [arXiv:1007.4937 [astro-ph.HE]].  
  
\bibitem{next} 
  C.~Bambi, A.~Cardenas-Avendano, T.~Dauser, J.~A.~Garcia and S.~Nampalliwar,
  %``Testing the Kerr black hole hypothesis using X-ray reflection spectroscopy,''
  arXiv:1607.00596 [gr-qc].  

\end{thebibliography}
\end{document}